\documentstyle[aps,epsfig]{revtex}
\def\b{\bar}
\def\d{\partial}
\def\D{\Delta}
\def\cD{{\cal D}}
\def\cK{{\cal K}}

\def\G{\Gamma}
\def\l{\lambda}

\def\m{\mu}
\def\n{\nu}

\def\q{\b q}

\def\t{\tau}

\def\~{\tilde}
\def\h{\eta}

\def\bY3{\bar Y_{,3}}
\def\Y3{Y_{,3}}
\def\z{\zeta}
\def\Z{{\b\zeta}}
\def\Y{{\bar Y}}
\def\cZ{{\bar Z}}
\def\`{\dot}
\def\be{\begin{equation}}
\def\ee{\end{equation}}
\def\bea{\begin{eqnarray}}
\def\eea{\end{eqnarray}}

\def\fn{\footnote}

\def\mn{{\mu\nu}}

\begin{document}
\title{Complex Kerr Geometry and Nonstationary Kerr
Solutions}

\author{Alexander Burinskii}
\address{Gravity Research Group, NSI Russian
Academy of Sciences, B. Tulskaya 52, 115191 Moscow, Russia}
\maketitle

\begin{abstract}
\noindent
In the frame of the Kerr-Schild approach, we consider
complex structure of the Kerr geometry which is determined by a
complex world line of a complex source.
The real Kerr geometry is represented as a real
slice of this complex structure.
The Kerr geometry is generalized to the nonstationary case when the
current geometry is determined by a retarded time and is
defined by a retarded-time construction via a given complex world line of
source.  A general exact solution corresponding to arbitrary
motion of a spinning source is obtained.  The acceleration of the
source is accompanied by a lightlike radiation along the principal null
congruence. It generalizes to the rotating case the known
Kinnersley class of "photon rocket" solutions.

PACS number(s): 04.20.Jb 04.20.-q 97.60.Lf 11.27.+d
\end{abstract}

\section{Introduction}

The complex representation of Kerr geometry, initiated by Newman
\cite{LinNew}, has been found to be useful in various problems
\cite{New02,BurStr,BurSup,BurMag}.
When considered in
the Newman-Penrose formalism \cite{NewPen,KraSte}, it allows one to
get a retarded-time description of the nonstationary Maxwell fields and
twisting algebraically special solutions of the linearized Einstein
equations.
Twisting solutions are represented in this approach as retarded-time
fields, which are similar to Lienard-Wiechard fields. However, they are
generated by a {\it complex} source moving along a {\it complex world line}
$x_0(\t)$ in complex Minkowski space-time $CM^4$.
The light cones emanating from the word line of a source usually play a
central role in the retarded-time constructions where the fields are
defined by the values of a retarded time. In the case of complex
world line, the corresponding light cone has to be complex which
complicates the retarded-time scheme.

In this paper, we use the Kerr-Schild approach to the complex representation
of Kerr geometry, which is based on the Kerr-Schild formalism and the Kerr
theorem\cite{DKS}. This approach has an advantage in this problem since
the Kerr-Schild form of metrics, $g^\mn=\eta ^\mn - 2h k^\m k^\n $,
contains the auxiliary Minkowski background $\eta^\mn$, which attaches an
exact meaning to the complex world line $x_0(\t)\in CM^4$ in the curved
Kerr-Schild backgrounds.  In addition, the Kerr theorem
allows us to get an explicit representation for the
metric, the principal null congruence (PNC),
formed by vector field $k^\m$, and the location of singularity and also
to express them in asymptotically flat Cartesian coordinates.

This approach  allows us to get a class of the exact nonstationary
generalizations of the Kerr solution which are determined by the
complex retarded-time construction.

The basic ideas of this approach were published in
\cite{IvBur1,Bur1,BKP,BurMag}.  In \cite{BurMag} this retarded-time
construction was applied to describe the boosted Kerr solution.
 However, application of this approach to accelerating twisting sources
encounters hard obstacles connected with the problem of a real slice.
In this paper we find the way to solve this problem and present the class
of exact nonstationary Kerr solutions.

For the non-rotating sources, such solutions  were obtained earlier
by Kinnersley \cite{Kin}. The Kinnersley solutions are radiative, and
by acceleration of the source they are accompanied by null radiation.
This property is present in our solutions also and leads to the necessity
of giving an interpretation to the origin of this radiation, which enforces
to return to the old problem of the source of the Kerr and Kerr-Newman
solutions.

The main peculiarity of the Kerr geometry is the twisting geodesic
and shear-free PNC.  Such congruences are determined in Minkowski space-time
(parametrized by the null Cartesian coordinates $u,v,\z,\Z$)
by the Kerr theorem \cite{KraSte,Pen,PenRin,CoxFla} via the solution
$Y(x)$ of the equation $F=0$, where $F(Y, \l _1, \l _2)$ is an arbitrary
analytic function of the projective twistor  variables \be Y,\qquad \lambda
_{1} = \z - Y v ,\qquad \lambda _{2} = u + Y \Z .  \ee In twistor notations
\cite{PenRin} these variables are defined as follows
$\{1, \ Y, \ \l _1, \  \l_2 \} = Z^\alpha / Z^0 $,
where
$Z^\alpha =\{ \mu ^A, \quad x^\nu \sigma _{\nu
\dot B B}  \mu ^{B} \}$.
One sees that $Y$ is the ratio of two components of
a spinor corresponding to the null direction which is tangent to a null ray
of
the PNC, while the coordinates $\{\l _1, \  \l _2 \}$ are connected with a
shift of this ray from the origin  and can be determined by any point
$x^\nu$
lying on this ray.  Therefore, these coordinates fix the position and
direction of a null ray in Minkowski space $M^4$, in accordance with
geometrical meaning of a null twistor, and the scalar function $Y(x)$
determines the null congruence as a field of null directions in $M^4$. The
geometrical meaning of the twistor coordinates is extended to $CM^4$, where
they fix complex null planes, and  the real null rays of congruence belong
to
the intersection of the complex conjugate null planes.

The retarded-time construction is usually based on the space-time links
provided by light cones.  When considering  the complex retarded-time
construction, we set up a link between  the generating PNC function $F$ and
the world line of a complex source. In the problem considered there
appears the obstacle that the real light cone does not have intersections
with complex world line.
In $CM^4$ light cones split on the families of `left' and `right'
complex null planes, which take over the role of the light cone
in the complex retarded-time scheme.  Each of the `left' null planes is
determined by the fixed values of the coordinates $\{ Y, \ \l _1, \  \l _2
\}
$ and represents a geometrical realization of the twistor \cite{PenRin}.

In Sec.II we recollect the basic properties of Kerr geometry and give a
nonformal treatment clarifying its complex structure.
A consequent treatment based on the Kerr-Schild formalism is started in
Sec.III.

In Appendix A we give the basic necessary relations of the Kerr-Schild
formalism, and in the Appendix B we give a proof of the Kerr theorem adapted
to the Kerr-Schild formalism. During the proof we obtain some relations that
are necessary for subsequent treatment of the real slice procedure described
in Sec.III.  It allows us to integrate the Einstein field equations, which
is
performed in Sec.IV.

\section{Complex structure of Kerr geometry and related
retarded-time construction}

\subsection{Appel source and main peculiarities of the real and complex Kerr
geometry}

{\bf The Kerr singular ring} is one of the most remarkable peculiarities of
the Kerr solution.  It is a branch line of space on two sheets: "negative"
and
"positive" where the fields change their signs and directions.
There exist the
Newton and Coulomb analogues of the Kerr solution possessing the Kerr
singular ring. This allows one to understand the origin of this ring
as well as the complex origin of the Kerr source.
The corresponding Coulomb solution was obtained by Appel still
in 1887 by a method of complex shift \cite{App}.  \par A point-like charge
$e$, placed on the complex z-axis $(x_0,y_0,z_0)= (0,0, ia)$, gives the real
Appel potential \begin{equation} \phi_a = Re \ e/{\tilde r} .
\end{equation}
Here $\tilde r$ is in fact the Kerr complex radial coordinate $\tilde r=
PZ^{-1}= r+ i a \cos\theta$, where $r$ and $\theta$ are the oblate
spheroidal
coordinates.  It may be expressed in the usual rectangular Cartesian
coordinates $x,y,z,t$  as
\begin{equation} \tilde r = [(x-x_0)^2 + (y-y_0)^2 +
(z-z_0)^2]^{1/2} = [x^2 + y^2 + (z-ia)^2]^{1/2}.  \end{equation}
The singular line of the solution
corresponds to $r=\cos\theta=0$, and it is seen that the Appel
potential $\phi_a$ is singular at the ring $z=0,\quad x^2+y^2=a^2$.  It was
shown that this ring is a branch line of space-time for two sheets, similar
to the properties of the Kerr singular ring.  Appel potential describes {\it
exactly} the e.m. field of the Kerr-Newman solution \cite{Bur0}.

If the Appel source is shifted to a complex point of space
$(x_o, y_o , z_o ) \rightarrow (0,0,ia)$, it can be considered as
a mysterious "particle" propagating along a {\it complex world-line}
$x_0^\mu (\tau)$ in $CM^4$ and parametrized by a complex time $\tau$.
The complex source of the Kerr-Newman solution
 has just the same origin \cite{LinNew,BurStr} and can be described by means
of a complex retarded-time construction for the Kerr geometry.  \fn{ The
objects described by the complex world-lines occupy an intermediate position
between particle and string.  Like a string they form  two-dimensional
surfaces or world-sheets in space-time \cite{BurStr,OogVaf}.  In many
respects this source is similar to the "mysterious" $N=2$ complex string of
superstring theory \cite{OogVaf}.}

{\bf The Kerr twisting PNC} is the second remarkable structure of the Kerr
geometry.  It is described by a vector field $k^\m$ which
determines the Kerr-Schild ansatz for the metric,
\be g_{\m\n} = \h_{\m\n} + 2 h k_{\m} k_{\n}, \label{ksa} \ee where
$ \h_\mn $ is an auxiliary Minkowski space-time and
\be h= \frac {mr-e^2/2} {r^2 + a^2 \cos^2 \theta}.
\ee
This is a remarkable simple form showing that all the complication
of the Kerr solution is included in the form of the field $k_\m (x)$
which is tangent to the Kerr PNC. This form shows also that the
metric is singular at $r=\cos\theta=0$, which are the focal points
of the oblate spheroidal coordinate system.

\begin{figure}[ht]
\centerline{\epsfig{figure=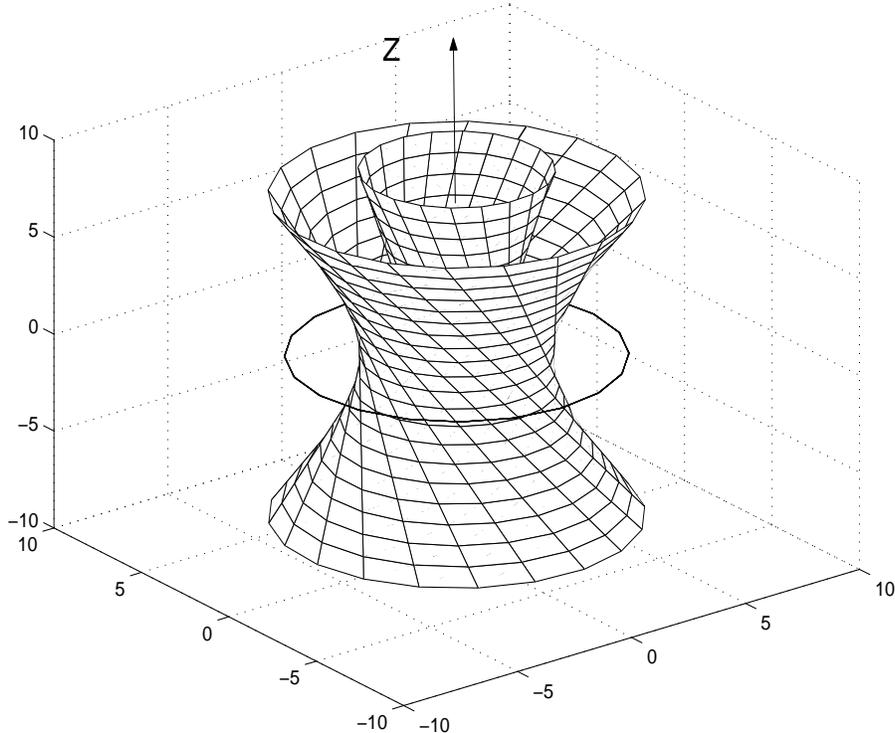,height=10cm,width=12cm}}
\caption{The Kerr singular ring and 3D section of the Kerr principal null
congruence. Singular ring is a branch line of space, and the PNC
propagates from the ``negative'' sheet of the Kerr space to ``positive ''
one,
covering the space-time twice. } \end{figure}

The field $k^\m$ is null with respect to $\h_\mn$ as well as with respect
to the metric $g_\mn$.
The Kerr singular ring and a part of the Kerr  PNC are shown in Fig.1.
The Kerr PNC consists of the linear generators of the surfaces $\theta
 =const$. The region shown in Fig.1  $z<0$ corresponds to a
``negative'' sheet of space ($r<0$)  where we set the null rays to be
``in''-going.

Twisting vortex of the null rays propagates through the singular ring
$r=\cos\theta=0$ and get ``out'' on the ``positive'' sheet of space ($z>0$).
Indeed, the Kerr congruence covers the space-time twice, and this picture
shows only the half of the PNC corresponding to $0>\theta>\pi /2$.  It has
to
be completed by the part for $\pi/2 >\theta>\pi$ which is described by
another system of the linear generators (having opposite twist). The two PNC
directions for each point $x^\m\in M^4$
correspond to the known twofoldedness of the Kerr geometry and
to the algebraically degenerate metrics of type D.

As it is explicitly seen from the expression for $h$, the Kerr gravitational
field has twovaluedness, $h(r) \ne h(-r)$, and so also do the other fields
on
the Kerr background. The oblate coordinate system turns out to be very
useful
since it also covers the space twice, for $r>0$ and $r<0$, with the
branch line on the Kerr singular ring.

The appearance of the twisting Kerr
congruence may be understood as a track  of the null planes of the
family of complex light cones emanating from the points of the complex
world line  $x_0^\m(\tau)$ \cite{BKP,BurStr} in the retarded-time
construction. It is very instructive to consider the following splitting
of the complex light cones.

\subsection{Splitting of the complex light cone}

The complex light cone $\cK$ with the vertex at some point $x_0$, written
in spinor form
\be {\cK }= \{x: x =
x^{\m}_{o}(\tau) + \psi ^{A}_{L} \sigma ^{\m}_{A \dot { A}} \tilde{\psi
}^{\dot{A}}_{R} \} \label{sclc} \ee
 may be split  into two families of null planes:
"left" $( \psi _{L}$ =const; $\tilde{\psi}_{R}$ -variable)
and "right"$( \tilde{\psi }_{R}$ =const; $\psi _{L}$ -variable).
These are the only two-dimensional planes that are wholly
contained in the complex null cone.  The rays of the principal null
congruence
of the Kerr geometry  are the tracks of these complex null planes (right or
left) on the real slice of Minkowski space.

The light cone equation in the Kerr-Schild metric coincides with the
corresponding equation in Minkowski space because the null directions
$k^\m $ are null in both metrics $g_\mn$ and $\eta _\mn $.
\par

In the null Cartesian coordinates
\bea
2^{1\over2}\z &=& x+iy ,\qquad 2^{1\over2} \Z = x-iy , \nonumber\\
2^{1\over2}u &=& z + t ,\qquad 2^{1\over2}v = z - t . \label{ncc}
\eea

the light cone equation has the form $\z\Z +uv =0$.
As usual, in a complex
extension to $CM^4$ the coordinates $u,v$ have to be considered as complex
and coordinates $\z$ and $\Z$ as independent.
On the real section, in $M^4$, coordinates $u$ and $v$ take the real values
and $\z$ and $\Z$  are complex conjugate.

The known splitting of the light cone on the complex null planes has a close
connection to spinors and twistors. By introducing the projective spinor
parameter $Y = \psi ^{1}/ \psi ^{0}$ the equation of complex light cone
with the vertex at point $x_0$,
\be (\z  -\z _0) (\Z -\Z_0) = - (u - u_0) ( v - v_0) , \ee
splits into two linear equations \footnote{It is a generalization of
the Veblen and Ruse construction \cite{Veb,Rus} which has been used for the
geometrical representation of spinors.}
\bea  \z  - \z _0 &=& Y (v - v_0 ), \\
-Y (\Z - \Z _0 ) &=& (u - u_0 ) \ ,
\label{split1}
\eea
describing the "left" complex null planes
(the  null  rays  in  the real space).  Another splitting
\bea  - \tilde {Y} (\z  - \z _0) &=&  (u - u_0 ), \\
(\Z - \Z_0 ) &=& \tilde{Y} (v - v_0 ) \ ,
\label{split2}
\eea
gives the "right" complex null planes.

Thus, the equations of the"left" null planes (\ref{split1}) can be written
in
terms of the three parameters
\be
Y,\qquad \lambda _{1} = \z - Y v,\qquad \lambda _{2} = u + Y \Z ,
\ee
as follows:
\be \lambda _{1} = \lambda ^0_{1} , \quad  \lambda _{2} = \lambda ^0_{2} \ ,
 \label{lnp} \ee
where
\be
\lambda ^0_{1} = \z _0 - Y v_0  , \qquad \lambda ^0_{2} = u_0 + Y\Z _0
\label{np} \ee
denote the values of these parameters at the point $x_0$.
These three parameters are the projective
twistor variables and very important for further consideration since
the Kerr theorem is
formulated in terms of these parameters. The above
splitting of the complex light cone equation shows their origin explicitly.
Note also that in the terms of the Kerr-Schild null tetrad
\begin{eqnarray}
e^1 &=& d \zeta - Y dv, \qquad  e^2 = d \bar\zeta -  \bar Y dv, \nonumber \\
e^3 &=&du + \bar Y d \zeta + Y d \bar\zeta - Y \bar Y dv, \nonumber\\
e^4 &=&dv + h e^3,\label{KSt}
\end{eqnarray}
the projective twistor parameters take the form
\bea
\l_1 &=& x^\m e^1_\mu ,
\nonumber\\
\l _2 &=&x^\m (e^3_\m-\Y e^1_\m ),
\label{ltwea}
\eea
and correspondingly
\bea
\l ^0_1 &=&x_0^\m e^1_\mu ,
\nonumber\\
\l ^0 _2 &=&x_0^\m (e^3_\m-\Y e^1_\m ).
\label{lambx}
\eea
The  ``left'' complex null planes of the complex light cone
at some point $x_0$ can be expressed in terms of the tetrad as follows
\be x_L = x_0(\tau) + \alpha e^1 + \beta e^3 \ ,  \label{L}
\ee
and the null plane equations (\ref{lnp}) follow then from Eq.(\ref{L}) and
the tetrad scalar products $e^{1\m} e^1 _\m =e^{1\m} e^3 _\m =
e^{3\m} e^3 _\m =0$.  Similar relations are valid also for the ``right''
null
planes with the replacement $e^1 \to e^2$.

The "left" null planes of the
complex light cones form a complex Kerr congruence which generates all the
rays of the principal null congruence on the real space.  The ray with polar
direction $\theta,\phi$ is the real track of the "left" plane corresponding
to $Y = \exp{i \phi} \tan (\theta /2) $ and belonging to the cone that is
placed at the point $x_0$ corresponding to $\sigma = a \cos (\theta).$ The
parameter $\sigma = Im \tau$ has a meaning only in the range $- a \leq
\sigma
\leq a$ where the cones have real slices.  Thus, the complex world line
$x_0(t,\sigma )$ represents a restricted two-dimensional surface or strip,
in
complex Minkowski space, and is really a world-sheet.\fn{It may  be
considered
as a complex open string with a Euclidean parametrization $\tau = t+i\sigma
,
\bar \t =t-i\sigma $, and with end points $x_0( t,\pm a)$
\cite{BurStr,OogVaf}.}

 The Kerr congruence arises as the real slice of the family of the
"left" null planes ($Y=const.$) of the complex light cones which vertices
lie on the complex world line $x_0(\tau)$.

The Kerr theorem can be linked to this retarded-time construction.

\section{Kerr theorem and the retarded-time construction}

\subsection{The Kerr theorem}

Traditional formulation of the Kerr theorem is following.

Any  geodesic and shear-free null congruence in
Minkowski space is defined by a function $Y(x)$ which is a solution of the
equation
\be         F  = 0 ,                            \label{(1.1)}\ee
where   $F (\l_1,\l_2,Y)$   is an arbitrary analytic
function of the projective twistor coordinates
\be
Y,\qquad \l_1 = \z - Y v, \qquad \l_2 =u + Y \Z .\label{(1.2)}
\ee
The congruence is determined then by the vector field
\be
 e^3 = du+ \Y d \z  + Y d \Z - Y \Y d v  = P k_\m dx^\m
\label{1.8}
\ee
in the null Cartesian coordinates
$u, \ v, \ \Z , \ \z $. \fn{The field $k_\m$ is a normalized form of
$e^3_\m$ with $k_\m Re \ \dot x_0^\m=1$.}

In the Kerr-Schild backgrounds
the Kerr theorem acquires a broader content \cite{DKS,IvBur1,BKP}.
It allows one to obtain the position of singular lines, caustics of the
PNC, as a solution of the system of equations
\be F=0;\quad d F / d Y =0 \ , \label{sing}\ee
and to determine  some important parameters of the corresponding solutions:
\be
\tilde r = - \quad d F / d Y , \label{tr}
\ee
and
\be
P = \d_{\l_1} F - \Y \d_{\l_2} F.  \label{PF}
\ee
The parameter $\tilde r$ characterizes a complex radial distance,
and for the stationary Kerr solution it is a typical complex
combination $\tilde r= r+ia \cos\theta$. The parameter $P$ is connected with
the boost of source.

The proof of the Kerr theorem in the extended  version adapted to the
Kerr-Schild formalism  is given in the Appendix B.
Some basic relations of the
Kerr-Schild formalism are given in Appendix A.

Working in $CM^4$ one has to consider $Y$ and $\bar Y$ functionally
independent, as well as the null coordinates $\z $ and  $\Z$. The
coordinates
$u,v$ and congruence turn out to be complex.  The
 corresponding complex null tetrad (\ref{Kt}) may be considered as a basis
of
$CM^4$.  The Kerr theorem determines in this case only the ``left" complex
structure - the function $F(Y)$. The real congruence appears as an
intersection with a complex conjugate ``right'' structure.

\subsection{Quadratic function F(Y) and interpretation of parameters.}
It is instructive to consider first the stationary case.
Stationary congruences having Kerr-like singularities contained in
a bounded region were considered in papers \cite{IvBur,Bur1,KerWil}.
It was shown that in this case the function $F$ must be at most quadratic in
$Y$, \be F \equiv a_0 +a_1 Y + a_2 Y^2 + (q Y + c) \l_1 - (p Y + \q) \l_2,
\label{FK}
\ee
where the coefficients $ c$ and $ p$ are real constants
and $a_0, a_1, a_2,  q, \q, $  are complex constants.
The Killing vector of the solution is determined as
\be
\hat K = c\d _u + \q \d _\z + q \d _\Z -p\d _v . \label{Knull}
\ee
Writing the function F in the form
\be
F = A
  Y^2 + B Y + C, \label{Fquadr}
\ee
one can find two solutions of the equation $F=0$ for the
function $Y(x)$
\be
 Y_{1,2} = (- B \pm \D )/2A, \label{Y12}
\ee
where $ \D = (B^2 - 4AC)^{1/2}.$

On the other hand from  (\ref{tr})
\be
\tilde r = - \d F /\d Y= -2AY -B,
\label{tr2}
\ee
and consequently
\be
\tilde r =PZ^{-1} = \mp \D.
\label{tr1}
\ee
These two roots reflect the known twofoldedness of the Kerr
geometry.  They correspond to two different directions of
congruence on positive and negative sheets of the Kerr space-time.
The expression (\ref{PF}) yields
\be
P=pY\Y  + \q \Y + qY +c \ . \label{Ppc}
\ee
\subsection{Link to the complex world line of the source.}
The stationary and boosted Kerr geometries are described
by
a  straight complex world line with a real 3-velocity $\vec v$ in $CM^4$:
\begin{equation}
x_0^\m (\t) = x_0^\m (0) + \xi^\m \t; \qquad \xi^\m = (1,\vec v)\ .
\label{dec}
\end{equation}
The gauge of the complex parameter $\t $ is chosen
in such a way that $Re \ \tau$ corresponds to the real time $t$.

The function $F$, quadratic in $Y$, can be expressed in this case
in the form \cite{IvBur,Bur1,BKP,BurMag} \begin{equation} F \equiv (\l_1 -
\l_1^0) \hat K\l_2 - (\l_2 -\l_2^0) \hat K\l_1 \ , \label{Fx0}
\end{equation}
where the twistor components $\l_1, \ \l_2 $ with zero indices denote their
values on the points of the complex worldline $ x_0 (\t)$, Eq.(\ref{np}),
and
$\hat K$ is a Killing vector of the solution \begin{equation} \hat K = \d
_\tau x_0^\m(\t) \d_\m = \xi^\m \d_\m \ . \label{hK} \end{equation}
Application of $\hat K$ to $\l_1$ and $\l_2$ yields the expressions
\bea
\hat K \l_1 &=& \d _\tau x_0^\m(\t) e^1_\mu ,
\nonumber\\
\hat K \l _2 &=&\d _\tau x_0^\m (e^3_\m-\Y e^1_\m ).
\label{cKll}
\eea
From Eq. (\ref{PF}) one obtains in this case
\be
P= \hat K \rho = \d _\tau x_0 ^\mu (\tau)e^3_\mu \ ,
\label{Prho}
\ee
where
\be
\rho= \l_2 + \Y \l_1 = x^\mu e^3_\mu \label{rho}.
\ee
Comparing Eqs. (\ref{Prho}) and  (\ref{Ppc}) one obtains the correspondence
in terms of $ \ p,\ c,\ q, \ \q $,
\be
\hat K\l _1= pY+\q, \qquad \hat K \l _2=qY +c,
\label{Kll}
\ee
which allows one to set the relation between the parameters
$p, c, q, \q$, and  $\xi ^\mu$,
showing that these parameters are connected with the boost of the source.

The complex initial position of the complex world line $x_0^\m(0)$  in Eq.
(\ref{dec}) gives six parameters for the solution, which are connected to
the coefficients $a_0, \ a_1 \ a_2 \ $. It can be decomposed as
$\vec x_0 (0) = \vec c + i\vec d$,
where $\vec c $ and $\vec d$ are real 3-vectors with
respect to the space O(3)-rotation.  The real part $\vec c$ defines the
 initial position of the source, and the imaginary part $\vec d$ defines the
value and direction of the angular momentum (or the size and orientation of
a singular ring).

It can be easily shown that in the rest frame, when $\vec v=0, \quad \vec d
=\vec d_0 $, the singular ring lies in the  orthogonal to $\vec d$ plane
and has a radius $a=\vert \vec d_0 \vert $.
 The corresponding angular momentum is $\vec J = m \vec d_0.$

\subsection{L-projection and complex retarded-time parameter.}

In the form (\ref{FK}) all the coefficients are constant while the
form (\ref{Fx0}) has an extra explicit linear dependence on $\tau$
via terms $ \l_1^0 (x_0(\t))$ and $ \l_2^0 (x_0(\t))$.
However,
this dependence is really absent. As a consequence of
the relations $ \l_1^0 (x_0(\t)) = \l_1^0 (x_0(0)) + \t \hat K \l_1,\quad
\l_2^0 (x_0(\t)) = \l_2^0 (x_0(0)) + \t \hat K \l_2 $, the terms
proportional
to $ \t$ cancel and these forms are equivalent.

Parameter $\t$ may be defined for each point
$ x$ of the Kerr space-time and plays the role of a complex retarded-time
parameter.  Its value for a given point $x$ may be defined by
L-projection, using the solution $Y(x) $ and forming the twistor parameters
$ \l_1,\quad \l_2 $ which fix a left null plane.

$L$-projection of the point $x$ on the complex world line $x_0(\t)$ is
determined by the condition
\be (\l_1-\l_1^0)|_L = 0,\qquad (\l_2-\l_2^0)|_L
=0 \ , \label{Lnp} \ee
where the
sign $|_L$ means that the points $x$ and $x_0(\t)$ are synchronized by the
left  null plane (\ref{L}),
\be x- x_0(\t_L) = \alpha e^1 + \beta e^3 . \ee
The condition (\ref{Lnp}) in representation
(\ref{lambx}) has the form
\be
 (x^\m -x_0^\m) e^1_\m |_L =0, \qquad  (x^\m -x_0^\m)
(e^3_\m - \Y e^1_\m)|_L=0,
\label{ll0}
\ee
which shows that the points $x^\m $
and $x_0^\m$ are connected by the left null plane spanned by the null
vectors $e^1$ and $e^3$.

This left null plane belongs simultaneously to the "in"-fold of the
light cone connected to the point $x$  and to the "out"-fold of the light
cone emanating from a point of the complex world line $x_0$.  The point of
intersection
of this plane with the complex world-line $x_0(\t)$ gives the value of the
"left" retarded time $\t_L$, which is in fact a complex scalar function on
the (complex) space-time $\t_L(x)$.

By using the null plane equation (\ref{Lnp}) one can  express $\D$ of
Eq. (\ref{tr1}) in the form
\be \Delta |_L= (u-u_0) \dot v_0 + (\z - \z_0 ) \dot
{\Z _0} + (\Z - \Z_0) \dot \z_0 + (v - v_0) \dot u_0 =\frac 12 \d_\t (x -
x_0
)^2 = \t _L -t + \vec v \vec R , \label{Dpm}\ee
where
\be \vec v = \dot {\vec
x_0}, \qquad \vec R = \vec x - \vec x_0. \label{3.23}\ee

It gives a retarded-advanced time equation
\be \t = t \mp \tilde r + \vec v  \vec R,
\label{ret-adv}\ee
and a simple expression for the solutions $Y(x)$:
\be Y_1 =  [ (u-u_0) \dot v_0 + (\z - \z_0) \dot{\Z}_0]
/ [ (v - v_0) \dot {\Z}_0 - (\Z -\Z_0) \dot v_0],
\ee
and
\be Y_2 =
[ (u-u_0) \dot \z _0 - (\z - \z_0) \dot{u}_0]/
[ (u-u_0) \dot v_0 + (\z - \z_0) \dot{\Z}_0] .
\ee

For the stationary Kerr solution $\tilde r=r+ia\cos\theta$,
and one sees that the second root $Y_2(x)$ corresponds to a transfer to
the negative sheet of the metric: $r\to -r; \quad \vec R \to -\vec R$, with
a
simultaneous complex conjugation  $ia \to -ia$.

Introducing the corresponding operations,
\be
P: r\to -r, \quad \vec R\to - \vec R \label{P},
\ee
\be
C: x_0 \to \bar x_0, \label{C}
\ee
and  also the transfer $"out" \to "in" $
\be
T: t-\tau \to \tau -t, \label{T}
\ee
one can see that the roots and corresponding Kerr congruences are
CPT-invariant.

\subsection{Nonstationary case.  Real slice.}

In the nonstationary case, this construction acquires new peculiarities.

i/ The coefficients of function $F$ turn out to be complex variables
depending
on the complex retarded-time parameter;

ii/ $\d _\tau x_0^\mu =\xi ^\mu $ can take complex values, which implies
complex values for the function $P$ and was an obstacle for obtaining the
real solutions in previous investigation \cite{BKP};

iii/ $K$ is no longer a Killing vector.

 To form the real slice of space-time, we have to consider, along with
the ``left'' complex structure generated by a ``left'' complex world line
$x_0$, parameter $Y$, and the left null planes, an independent ``right''
structure with  the``right'' complex world line $\bar x_0$, parameter $\Y$,
and the right null planes, spanned by $e^2$ and $e^3$.
These structures can be
considered as functionally independent in $CM^4$, but they have to be
 complex conjugate on the real slice of  space-time.

First, note that for a real point of space-time $x$ and for the
corresponding
real null direction $e^3$,  the values of the function
\be \rho (x) = x^\m e^3_\m(x)
\label{rhonst}
\ee
are real.
Next, one can determine the
values of $\rho$ at the points of the left and right complex world lines
$x_0^\mu$ and $\bar x_0 ^\mu$ by L- and R-projections
\be \rho _L (x_0) =
x_0^\m e^3_\m (x)|_L \label{rhoL}\ee
and
\be \rho _R(\bar x_0) = \bar x_0^\m
e^3_\m (x)|_R .\label{rhoR}\ee
For the "right" complex structure, the points $x$ and $\bar x_0(\bar \t)$
are to be synchronized by the right null plane
$ x - \bar x_0(\bar \t _R) = \alpha e^2 + \beta e^3 $.
As a consequence of the
conditions $e^{1\mu} e^3_\mu=e^{3\mu} e^3_\mu =0$, we obtain
\be \rho _L(x_0) = x_0^\m e^3_\m (x)|_L = \rho (x).
\label{rho0}\ee
So long as the parameter
$\rho (x) $ is real, the parameter $\rho _L (x_0)$ will be real, too.
Similarly,
\be \rho _R(\bar x_0) = {\bar x_0}^\m e^3_\m (x)|_R = \rho (x),
\ee and consequently, \be \rho _L(x_0) = \rho (x) = \rho _R (\bar
x_0).
\label{(4.14)}\ee
By using  Eqs.(\ref{lambx}) and (\ref{rhonst}) one obtains
\be \rho = \l_2 + \Y
\l_1.  \label{rhotw} \ee
Since the L-projection (\ref{Lnp}) determines the  values of the
left retarded-time parameter $\tau _L= (t_0 + i\sigma)|_L$,
the real function $\rho$ acquires an extra dependence on the
retarded-time parameter $\tau _L$.
 It should be noted that
the real and imaginary parts of $\tau |_L$
are not independent because of the constraint caused by
L-projection.

It means that the real functions $\rho $ and
 $\rho _0 $ turns out to be functions of the real retarded-time parameter
 $t_0= Re  \  \tau _L$, while
 $\l^0_1$ and $\l^0_2$ can also depend on
$\sigma $.

These parameters are constant on the
left null planes, which yields the relations
\be
(\sigma |_L),_2 =(\sigma |_L),_4 = 0 , \quad
(t_0|_L),_2 =(t_0|_L),_4 = 0 \ . \label{st02}
\ee
Similar to  the stationary case considered above,
we shall restrict function $F$ by the expression quadratic in $Y$
\be
F \equiv (\l_1 - \l_1^0) K_2 - (\l_2 -\l_2^0) K_1  ,
\label{Fnst}
\ee
where the functions $K_1$ and $K_2$ are linear in $Y$ and depend on the
retarded-time $t_0$. It has to
lead to the form (\ref{FK}) with the coefficients depending on the
retarded-time.

Let us assumme that the relation $F(Y,t_0)=0$ holds for the retarded-time
evolution $\d F / \d {t_0}|_L=0.$
It yields
\be \frac {\d F}{\d {t_0}}= K_1
\d_{t_0} \l_2^0 - K_2 \d_{t_0} \l_1^0  +
(\l_1 -\l_1^0) \d_{t_0} K_2 - (\l_2-\l_2^0) \d_{t_0} K_1 =0.
\label{dFdt0} \ee
As a consequence of L-projection the last two terms cancel and one obtains
\be ({\d F}/{\d {t_0}})|_L= (K_1
\d_{t_0} \l_2^0 - K_2 \d_{t_0} \l_1^0)|_L =0,  \label{dFK12} \ee
that is provided by
\be
 K_1 (t_0) =\d_{t_0} \l_1^0 ,\qquad K_2 (t_0) =\d_{t_0} \l_2^0 \ .
\label{K12} \ee
In tetrad representation (\ref{lambx}) it takes the form
\be
 K_1 =\d _{t_0} x_0^\m e^1 _\m ,\qquad
K_2 =\d _{t_0} x_0^\m (e^3 _\m -\Y e^1 _\m).
\label{K12t} \ee
As a consequence of the relation (\ref{PF}), one obtains
\be
P=\Y K_1 +K_2 ,
\ee
which yields for the function $P$ the real expression
\be
P=\d _{t_0} (x_0^\m e^3_\m)|_L = \d _{t_0} \rho _L \ .
\label{Pnst}
\ee
It is seen that $\rho(t_0)=\rho _L(t_0)$ plays
the role of a potential for $P$, similarly to some nonstationary solutions
presented in \cite{KraSte}.

It seems that the extra dependence of the function $F$ on the nonanalytic
retarded-time parameters $t_0$  contradicts the Kerr theorem;
however, the nonanalytic part disappears  and analytic
dependence on $Y,\quad l_1, \quad l_2$ is reconstructed by L-projection.
This is explicitly seen for the quadratic form (\ref{Fnst})
with coefficients given by Eq.(\ref{K12}).
Indeed, direct differentiation of this
form yields the expression
\be dY = ae^1 +be^3 +cdt_0 \ , \label{dYabc} \ee
where $c=(\l_1 -\l_1^0) \d_{t_0} K_2 - (\l_2-\l_2^0) \d_{t_0} K_1$.  By
L-projection one has $c|_L=0$ and the nonanalytic term $c dt_0$ cancels.
Therefore, the differential of the function $Y(x)$ by L-projection satisfies
the geodesic and shear free conditions (\ref{2.1}) provided by the Kerr
theorem.  Note that all the {\it real} retarded-time derivatives on the real
space-time are nonanalytic and have to involve the conjugate right complex
structure. In particular, the expressions (\ref{K12t}) acquire the form \be
 K_1 =e^1 _\m Re \ \dot x_0^\m  ,\qquad
K_2 =(e^3 _\m -\Y e^1 _\m) Re \ \dot x_0^\m  ,
\label{K12dot} \ee
where $\dot x_0^\m = \d _{t_0} x_0^\m \ .$

\section{Solution of the field equations}
\bigskip
We are now able to obtain a general class of accelerating radiating
solutions representing arbitrary nonstationary generalization of the
Kerr solution.
For simplicity, we shall assume that there is no electromagnetic field.
As in the Kinnersley case, the null radiation is described by an
incoherent flow of the lightlike particles in $e^3$ direction.
 The solution of the field equations is similar to the treatment given
for the Kerr-Schild form of metric in \cite{DKS} \fn{There are no changes up
to Eq.  (5.50) of this work.}
\par
In particular, we have
\be R_{24} =R_{22} =R_{44}=0.  \label{(5.1)}\ee
If the electromagnetic field is zero we also have
 \be R_{12} =R_{34}=0.   \label{2}\ee
The equation
\be h,_{44} + 2(Z + \cZ)h,_4 + 2Z \cZ h =0,
\label{(5.4)}\ee
which  follows from (\ref{2}), admits the
solutions \be h= M(Z+ \cZ)/2  ,\label{(5.5)}\ee
where $M$ is a real function,
obeying the conditions $ M,_4=0 $.
\par
  Next, the equation \be R_{23} = 0 , \label{(5.6)}\ee  acquires the form
\be M,_2 - 3 Z^{-1} \cZ Y,_3 M  = 0.\label{7}\ee

  The last gravitational field equation
$R_{33} = -P_{33}\equiv - \kappa T_{33}$ takes the form
\be \cD M
= Z^{-1} \cZ ^{-1} P_{33}/2  ,
\label{last}\ee
where
\be
\cD=\d _3 - Z^{-1} Y,_3 \d_1 - \cZ ^{-1} \Y ,_3 \d_2   \ , \label{cD}
\ee
which  corresponds to the null radiation in the form
$\kappa T_\mn = P_{33}e^3_{\m} e^3_{\n}$.
\par
To integrate Eq.(\ref{7}) we use the relation (\ref{2.11}) of corollary 1
and obtain the equation
\be
(\log MP^3),_2=0
\ee
which has the general solution
\be M= m/P^3,
\label{11}\ee
where \be m,_4 = m,_2 =0. \label{(5.12)}\ee
On the real slice functions $m$ and $P$ depend on the retarded-time $t_0$.
The action of operator $\cD$ on the variables
$Y, \bar Y $ and $ \rho$ is
\be \cD Y = \cD \bar Y = 0,\qquad
\cD \rho =1 \ . \label{16}\ee
From these relations and Eq. (\ref{Pnst}) we have
$\cD \rho = \d \rho / \d t_0 \cD t_0  = P\cD t_0 =1 $,
which yields
\be
\cD t_0 = P^{-1} .
\ee
Since $ M$ is also a function of $Y, \bar Y$, and $ t_0$, the last equation
(\ref{last}) takes the form
\be \partial _{t_0} M = P  Z^{-1} \cZ ^{-1}
P_{33}/2 . \label{5.17}\ee

It is not really a field equation but a
definition of the stress-energy tensor $\kappa T_\m^\n = P_{33}e^3_\m
e^{3\n}$ corresponding to the null radiation.
Substituting Eq.(\ref{11}) one
obtains two terms:
\be P_{33} = \frac 1 {P^2 |\tilde r|^2}  [-6m(\partial
_{t_0} \log P) + 2 \partial _{t_0} m].  \label{P33}\ee
The first term, proportional to $\partial _{t_0} \log P$,
is connected with the acceleration. The second term,
proportional to $\partial_{t_0} m $,
describes the loss of mass by radiation corresponding to the
Vaidia ``shining star'' solution \cite{KraSte,VaiPat,FroKhl}.

The resulting metric
   has the form \be g_{\m\n} = \h_{\m\n} + (m/P^3)(Z +\bar Z) e^3_{\m}
e^3_{\n}.  \label{metr}\ee
Normalizing $e^3$ by introducing $k^\m = e^{3 \m}/P$,
one has
\be {\dot x}_0^\m k_\m =1,
\label{norm} \ee
and using $\tilde r = PZ^{-1}$ we simplify the expressions for metric and
stress-energy tensor.

Let us summarize the solutions we have obtained.
The metric is
\be g_{\m\n} =\h_{\m\n} + m({\tilde r}^{-1} + \bar{\tilde r}^{-1})
k_{\m} k_{\n}, \label{f1}\ee
and the radiation is
\be
\kappa T_\m^\n = \Phi k_\m k^\n, \label{rad}
\ee
where
 \be \Phi = \frac 1 {|\tilde r|^2}  [-6m(\partial _{t_0} \log P) + 2
 \partial _{t_0} m].  \label{Phi}\ee
Vector field $k_\m$ is defined by
 \be k_\m d x^\m= e^3/P
 = P^{-1}(du+ \Y d \z + Y d \Z - Y \Y d v ). \label{f10} \ee
Function $Y(x)$ is given in terms of the coordinates
$u,v,\z,\Z$, by equation  $F(Y) =AY^2 +BY +C=0$, where the coefficients
$A,B,C$ are determined by decomposition of function
\be
F \equiv (\l_1 - \l_1^0) K_2 - (\l_2 -\l_2^0) K_1  ,
\label{f3}
\ee
and $P$ is given by
 \be P (t_0) =K_2 +\Y K_1 = e^3 _\m Re \ \dot x_0^\m  \ ,
\label{f9} \ee
where
\be \lambda _{1} = \z - Y v ,
\qquad \lambda _{2} = u + Y \Z ,
\label{f4} \ee
and
\be \lambda ^0_{1} = \z _0 - Y v_0  ,
\qquad \lambda ^0_{2} = u_0 + Y\Z _0
\label{f5} \ee
are the values of these variables on the complex world line,
and
\be
 K_1 =e^1 _\m Re \ \dot x_0^\m  ,\qquad
K_2 =(e^3 _\m -\Y e^1 _\m)Re \ \dot x_0^\m .
\label{f7} \ee
The complex radial coordinate is given by
\be \tilde r =- dF/dY = -2AY -B.
\label{f2}\ee
The coefficients $A, B, C$, the functions $K_1$ and $K_2$ and the parameters
of the function $P$ are determined by a given complex world line
$x_0^\m (\t) =\{u_0(\t), v_0(\t), \z_0(\t), \Z_0(\t) \}$ and have a current
dependence on the retarded time $\t|_L$ which is determined
by L-projection on the given complex
world line as a root of the left null plane equation
\be \lambda _{1} =
\lambda ^0_{1} \ , \quad \lambda _{2} = \lambda ^0_{2} \ .  \label{f6} \ee

Solution of these equations has to be performed for all points of
space-time in the region of interest. This is a nonlinear
problem with many unknown functions. In the general case it needs
a large body of numerical computations with subsequent
iterative refinement.
For the beginning of the iterative process a starting `point' is necessary,
which gives an initial approximation.
To obtain it the analytical solutions in the local regions of
short distances
$|\tilde r|\sim |t -\t| <<\dot x_0^\m k_\m/\ddot x_0^\m k_\m$,
can be used since all
the unknown parameters (for exclusion of radiation) are determined by the
first derivative of the complex world line and the nonlinearity caused by
acceleration  is negligible here. Having at hand the initial
field $Y(x)$, one can use the following iterative loop scheme of
computation:

\be x\to Y(x) \to \{\l_1, \l_2, \rho \} \to \{\t, x_0, \l^0_1, \l^0_2 \}
\to \{P, K_1, K_2 \} \to F(Y) \to \{ Y(x), e^{3} \} \to cycle \ loop \ .
\ee
This iterative procedure is needed to refine the data and carry out
a progressive extension of the region.
The obtained local parameters
$Y, K_1, K_2, \l^0_1, \l^0_2, A, B, C, F, P, \t|_L $
can be immediately extended {\it along the rays} of PNC  from the short
distances to large ones, which allows one to reduce considerably the
necessary body of computations.

The following example is instructive since it shows that some of
parameters can be determined  analytically; however, an essential
nonlinearity is retained which demands the numerical computations.

{\bf Example}

Let us consider circular motion of source in (x,y)-plane with the
direction of angular momentum $\vec J=m \vec a$ along z-axis.
This example is interesting for astrophysical applications and as a model
of circular motion of polarized spinning particles in accelerators.
For the both cases one can assume $|a|<|b|$.
The corresponding complex world line will be
\be x^\m _0 (\t) = (\t, \ b \cos
\omega t_0, \  b \sin \omega t_0, \ ia \ ), \qquad \t = t_0 +i\sigma .
\label{circ}\ee
In null coordinates $u,v,\z,\Z$ it takes the form \be x _0
(\t) = 2^{-1/2} (ia +\t, \ ia- \t , \  b e^{i\omega t_0}, \ b e^{-i\omega
t_0}
\ ) , \label{0x0}\ee
and we have \be \dot x _0 (\t) = 2^{-1/2} (1, \ - 1 , \
ib \omega e^{i\omega t_0}, \ -ib \omega e^{-i\omega t_0} \ ) .
\label{0dx0}\ee

The expression for $\rho _0$ will be
 \be \rho _0 = 2^{-1/2} \lbrack \t (1+Y\Y)
+ b (\Y e^{i\omega t_0}  +Y  e^{-i\omega t_0})  +ia (1-Y\Y) \rbrack .
\label{0rho0}\ee
On the left null plane it has to be real that leads to
\be
\sigma |_L   = a(1-Y\Y)/(1+Y\Y),
\label{sigL}\ee
and
\be
\rho _0 |_L= 2^{-1/2} \lbrack t_0 (1+Y\Y) +
b (\Y  e^{i\omega t_0}  +Y  e^{-i\omega t_0}) \rbrack,
\label{0rhoL}\ee
and yields
\be
P= \dot \rho _0 |_L= 2^{-1/2} ( 1+Y\Y) +
i 2^{-1/2}\omega b (\Y  e^{i\omega t_0}  -Y  e^{-i\omega t_0} ).
\label{0P}\ee
One can also obtain
\be
\l ^0 _1 = 2^{-1/2}
[b  e^{i\omega t_0}  + Y (\t _L-ia) ], \quad
\l ^0 _2 =  2^{-1/2}
[Y b  e^{-i\omega t_0}  +  (\t _L+ia)] ,
\label{0lamb}\ee
and
\be
K _1 = 2^{-1/2}[ib \omega e^{i\omega t_0}  + Y ] , \quad
K_2 = 2^{-1/2}[ 1- Y ib \omega e^{-i\omega t_0}].
\label{0K12}\ee
Coefficients of the function $F= AY^2 +BY +C$ take the form
\be
A =2^{-1/2} \{ ib \omega e^{-i\omega t_0} [v- 2^{-1/2}(ia -\t _L)]-
[\Z -2^{-1/2} b e^{-i\omega t_0} ] \},
\label{0A}\ee
\be
B =-2^{-1/2} \{ ib \omega e^{-i\omega t_0} [\z -2^{-1/2} b e^{-i\omega
t_0} ]
+ ib \omega e^{i\omega t_0} [\Z -2^{-1/2} b e^{i\omega t_0} ] +
(u+v -2^{1/2} ia)\} ,
\label{0B}\ee
\be
C =2^{-1/2} \{- ib \omega e^{i\omega t_0} [u- 2^{-1/2}(ia +\t _L)] +
[\z -2^{-1/2} b e^{i\omega t_0} ] \}.
\label{0C}\ee

However,  since these coefficients depend on the parameter
$\sigma = Im \ \t _L $, which is determined by L-projection as a
function of $Y$, the equation $F=0$ turns out to be nonlinear.
The iterative procedure is necessary for its solution.
The dependence of $A,B,C$ on $\sigma$ has the
factor $b\omega$.  The case $b\omega <<1$ corresponds to
nonrelativistic motion. The dependence on $\sigma$ is weak when
$b\omega << |Y|\sim 1$, but
grows when $b\omega \sim |Y|$ or $b\omega \sim |1/Y|$.
Neglecting $\sigma$ in the equation $F=0$, one can obtain the analytical
solution $Y(x)$, which can be used as a first approximation.
Note also that in the distant zone the
role of rotation parameter $a$ becomes weak and the simpler Kinnersley
solution can be used for correction of the parameters.

{\it Transfer to the Kinnersley solutions.}

For the twist-free, nonrotating Kinnersley's case the world line is real,
$ Im \ x_0 = 0 $,
and the radial distances $ \tilde r = \bar {\tilde r} =r$
and the `right' and `left' retarded-time parameters coincide
$\t _L= \t _R=t_0$.
The retarded-time equation following from Eq.(\ref{ret-adv})
can be represented in the form
\be
\tilde r = - (t -t_0) + \vec v  \vec R = (x^\m - x_0^\m) \dot x_{0\m}.
\label{ret}\ee
It turns out to be real, and in the terms of the Kinnersley
parameters  $\sigma ^\m= x^\m -x_0 ^\m$
it yields the relation
\be
\tilde r=r=\sigma ^\m \dot x_{0\m} .
\label{realret}\ee

On the other hand, the real null vectors
  $\sigma ^\m=x^\m -x_0 ^\m$ are proportional to $e^{3\m}$,
and taking into account Eq.(\ref{f9}) we have
$r=\sigma ^\m \dot x_{0\m}= \beta e^{3\m}\dot x_{0\m} = \beta P $,
which yields $\beta =r/P$ and
$\sigma ^\m = r e^{3\m}/P = r k^\m$.
This relation shows that our PNC field $k^\m$ coincides with the Kinnersley
definition of the PNC, $ k^\m=\sigma ^\m /r$, in terms of which
 \be \d _{t_0} \log P = \dot P/P =\ddot x_0^\m k_\m \ ,\label{ddot}
\ee
and the metric (\ref{f1}) and radiation  (\ref{rad}),  (\ref{Phi})
take the Kinnersley form  \cite{Kin}
\be g_{\m\n} =\h_{\m\n} +
\frac{2m(t_0)}{r} (\sigma_{\m}/r) (\sigma_{\n}/r), \label{kin}\ee
\be
\kappa T_\m^\n = \Phi k_\m k^\n ,
\qquad \Phi= \frac 1 {r^2}  [-6m(\ddot x_0^\m k_\m) + 2 \dot m].
\label{RPhi}\ee

\bigskip
\section{Conclusion}

The complex retarded-time construction considered permits us to obtain
a class of nonstationary rotating solutions generated by a complex source
moving along an arbitrary given complex world line.

These solutions represent a natural generalization of the Kinnersley
class of solutions to the rotating case.  The Kerr-Schild approach
allows one to get exact expressions for the metric, coordinate system, the
PNC, and the positions of singularity for arbitrary motion of a rotating
source.

The  solutions obtained represent a natural generalization of the black
hole solutions, and if $m^2<e^2 +a^2$ they have horizons.
However, since the solutions are radiative, the usual black-hole
interpretation can meet objections, and an additional treatment of this case
is necessary, which we intend to do elsewhere.  The solutions can
find application for modelling the behavior of spinning astrophysical
objects by acceleration and relativistic boosts.  They are also
interesting for investigation of the relativistic gravitational fields by
particle scattering in ultrarelativistic regimes \cite{BurMag}.

By $a^2+e^2 >m^2$ the horizons disappear and there is a naked singular ring.
This case has attracted attention as a model of a spinning particle
\cite{IvBur1,Bur1,Bur0,IvBur,Car,Isr,Lop}. In \cite{IvBur1,Bur1,IvBur}
the Kerr singular ring was considered as a closed
 relativistic string forming the source of spinning particle \cite{Bur0}.
The nonstationary Kerr solutions presented allows one to describe
excitations
of this string. In this case, the ``negative'' sheet of the Kerr space has
to
be considered as a sheet of advanced fields belonging to the field of vacuum
fluctuations, and thus the e.m. radiation must belong to the zero point
field.
The outgoing vortex of the null radiation appears as a result of the
resonance of the vacuum field on this relativistic string.  In this case,
the
energy-momentum tensor has to be regularized on the classical level by the
known procedure \cite{deWit}
\be T_{reg}^\mn = \ :T^\mn: \ \equiv T^\mn -<0|T^\mn|0>, \ee
which has to satisfy the condition
$\nabla _\m T_{reg}^\mn=0$.
It corresponds exactly to a subtraction of this radiation, leading to
$T_{reg}^\mn =0$ \cite{IvBur1,Bur1}.  On the quantum level this procedure is
equivalent to the postulate on the absence of radiation for oscillating
strings.  This stringy interpretation of the Kerr source  will be considered
elsewhere.

The class of solutions presented can easily be generalized to
the Kerr-Newman solution and to the sources radiating electrical
charges.

Some other known generalizations of the Kerr solution,
such as the Kerr-Sen solution to low energy
string theory \cite{Sen}, the solution to broken $N=2$ supergravity
\cite{BurSup}, and regular rotating particlelike objects built on the
base of Kerr-Newman solution \cite{BEHM}, retain the form
and the geodesic and shear-free properties of the Kerr PNC.
This means that the
Kerr theorem is also valid for these solutions, and that they can
also be generalized to the nonstationary radiating case.

\section*{Appendix A: Basic relations of the Kerr-Schild formalism}
Following the notations of Ref.\cite{DKS},  the Kerr-Schild null tetrad
 $e^a =e^a_\m dx^\m $ is determined by the relations
\begin{eqnarray}
e^1 &=& d \zeta - Y dv, \qquad  e^2 = d \bar\zeta -  \bar Y dv, \nonumber \\
e^3 &=&du + \bar Y d \zeta + Y d \bar\zeta - Y \bar Y dv, \nonumber\\
e^4 &=&dv + h e^3,\label{Kt}
\end{eqnarray}
and
\be g_{ab}= e_a^\m e_{b\m} = \left(
\begin{array}{cccc} 0&1&0&0 \\ 1&0&0&0 \\ 0&0&0&1 \\ 0&0&1&0 \end{array}
\right). \label{gab} \ee
The vectors
$e^3, e^4$ are real, and $ e^1, e^2 $ are complex conjugate.

 The Ricci rotation coefficients are given by
\be \G ^a_{bc} = - \quad e^a_{\m;\n} e_b^\m e_c^\n.  \label{(1.4)}
\ee
 The PNC have the $e^3$  direction as tangent.  It will be geodesic if and
only if $\G_{424} = 0$ and shear free if and only if $\G_{422} = 0$.  The
 corresponding complex conjugate terms are $\G_{414} = 0$ and $\G_{411} =
0$.

The inverse (dual) tetrad has the form
 \bea
  \d_1 &=& \d_\z  - \Y \d_u ,
\nonumber\\
\d_2 &=&  \d_\Z - Y \d_u ,
\nonumber\\
 \d_3 &=&  \d_u - h \d_4  ,
\nonumber\\
 \d_4 &=&  \d_v + Y \d_\z + \Y \d_\Z - Y  \Y \d_u ,  \label{1.10}
\eea
where $\d _a \equiv ,_a \equiv e_a^\m \d ,_\mu $.

The parameter $Z=Y,_1 =\rho +i \omega$ is a complex expansion of the
congruence, $\rho=expansion$ and $\omega = rotation$. $Z$ is connected to
the
complex radial distance $\tilde r$ by the relation
\be \tilde r =PZ^{-1}.
\label{trZ} \ee

It was shown in \cite{DKS} that the connection forms in Kerr-Schild metrics
 are
\be \G_{42} = \G_{42a} e^a  = - d Y - h Y,_4
e^4 .  \label{1.11}
\ee
The congruence  $e^3 $ is geodesic if $ \G_{424} =
-Y,_4 (1-h) = 0, $ and is shear free if $ \G_{422} = -Y,_2 = 0.$
Thus,  the function $ Y (x)$ with the conditions
\be
Y,_2 = Y,_4 = 0  \label{1.12}
\ee
defines a shear-free and geodesic congruence.

\section*{Appendix B: Proof of the Kerr Theorem and two corollaries.}
The proof given bellow  of the Kerr theorem follows to the
general scheme sketched in \cite{DKS}.

{\it Proof.}
The  differential of
the function $Y$  in the case of $Y,_2 =  Y,_4 = 0 $ has the form
\be
d Y = Y,_a e^a  = Y,_1 e^1 + Y,_3 e^3.
  \label{2.1}\ee
As the first step we work out the form of $Y,_3$. By using relations
(\ref{1.10}) and their commutators we find
 \be   Z,_2 = (Z - \bar Z) Y,_3.
  \label{2.2}\ee
Straightforward differentiation of $Y,_3$ gives the equation
\be
Y,_{32} = ( Y,_3) ^2   ,
\label{2.3}\ee
and by using Eqs.(\ref{2.2}) and (\ref{2.3}) we obtain the equation
\be
(Z^{-1} Y,_3),_2 = \bar Z ( Z^{-1} Y,_3 )^2.
\label{2.4}\ee

This is a first-order differential equation for the function
$Z^{-1} Y,_3$.
Its general solution can be obtained by the substitution $x=Z(Y,_3)^{-1}$
and has the form
\be
Y,_3 = Z ( \phi - \bar Y) ^{-1},
\label{2.5}\ee
where $\phi$ is an arbitrary solution of the equation $ \phi,_2 =0$.
Analogously, by using the relation $Y,_{34}=-Z Y,_3$ one gets
$\phi,_4=0$; therefore $\phi$ may be an arbitrary function satisfying
\be
\phi,_2=\phi,_4 =0.
\label{2.6}\ee
One can also mention that the three projective twistor coordinates
$ \l_1 = \z - Y v, \qquad \l_2 =u + Y \Z,$ and $ Y$
satisfy similar relations $ (.),_2=(.),_4 =0$. Since the surface
$\phi=const$ forms a sub-manifolds of  $CM^4$ that has the complex dimension
3, an arbitrary function $\phi$ satisfying Eq.(\ref{2.6}) may be presented
as function of three projective twistor coordinates $\phi=\phi(Y, \l_1,
\l_2)$.  Now we can substitute $Y,_3$ in Eq.(\ref{2.1}), which implies
\be Z^{-1}
(\Y - \phi) d Y = \phi (d \z -Y d v) + (du + Y d \Z).  \label{2.7}\ee If an
arbitrary analytic function $F (Y,\l_1,\l_2)$ is given, then differentiating
the equation $F (Y,\l_1,\l_2)=0$ and comparing the result with
 Eq.(\ref{2.7}), we find that
 \be PZ^{-1}= - \quad d F / d Y  , \qquad P = \d_{\l_1} F - \Y
 \d_{\l_2} F, \label{2.8}\ee
 where the function $P$ can also be defined as
 \be P = (\phi - \b Y)\d_{\l_2} F.
\label{2.9}\ee

{\it Corollary 1.}
The following useful relations are valid:
\be
\bar Z Z^{-1} Y,_3=- (\log P),_2 \ , \quad P,_4=0 .
\label{2.11}\ee

{\it Proof.}
So long as $\d_2 \d_{\l_2}F =0$, one sees that
\be
(\log P),_2 = -\bar Z (\phi -\Y)^{-1};
  \label{2.10}\ee
then Eq.(\ref{2.5}) leads to first equality of Eq.(\ref{2.11}).
The relation $P,_4=0$ follows from Eq.(\ref{2.8}) and the properties of
the twistor components $Y,_4= (\l_1),_4 =  (\l_2),_4 =0$.

{\it Corollary 2.}
Singular region of the congruence, where the complex divergence
$Z$ blows up, is defined by the system of equations (\ref{sing}).

\section*{Acknowledgments}
We are thankful to G. Alekseev and M. Demianski for very useful discussions.

\end{document}